\documentclass[prb,aps,preprint,superscriptaddress,showpacs]{revtex4-1}

\usepackage{graphicx}
\usepackage{dcolumn}
\usepackage{bm}

\begin{document}
\title{Improved room-temperature-selectivity between Nd and Fe in Nd recovery from Nd-Fe-B magnet}

\author{Y. Kataoka}
\affiliation{Department of Electrical Engineering, Faculty of Engineering, Fukuoka Institute of Technology, 3-30-1 Wajiro-higashi, Higashi-ku, Fukuoka 811-0295, Japan}
\author{T. Ono}
\affiliation{Physonit Inc., 6-10 Minami-Horikawa, Kaita Aki, Hiroshima 736-0044, Japan}
\author{M. Tsubota}
\affiliation{Physonit Inc., 6-10 Minami-Horikawa, Kaita Aki, Hiroshima 736-0044, Japan}
\author{J. Kitagawa}
\affiliation{Department of Electrical Engineering, Faculty of Engineering, Fukuoka Institute of Technology, 3-30-1 Wajiro-higashi, Higashi-ku, Fukuoka 811-0295, Japan}

\date{\today}

\begin{abstract}
The sustainable society requires the recycling of rare metals. 
Rare earth Nd is one of rare metals, accompanying huge consumption especially in Nd-Fe-B magnets.
Although the wet process using acid is in practical use in the in-plant recycle of sludge, higher selectivity between Nd and Fe at room temperature is desired.
We have proposed a pretreatment of corrosion before the dissolution into HCl and the oxalic acid precipitation.
The corrosion produces $\gamma$-FeOOH and a Nd hydroxide, which have high selectivity for HCl solution at room temperature.
Nd can be recovered as Mn$_{2}$O$_{3}$-type Nd$_{2}$O$_{3}$.
The estimated recovery-ratio of Nd reaches to 97\%.
\end{abstract}



\maketitle

\clearpage

\section{Introduction}
Rare-earth is widely consumed in glass industry, catalysts, Nd magnets and so on\cite{Goonan:report2011}.
Especially the demand of Nd-Fe-B magnets is rapidly growing, viewing from sustainable and/or low-carbon society, in motors of electric vehicle and wind turbine et al.
Major product-country of rare earths is now China, often controlling the export quota.
Therefore, the other countries urgently work through the recycle of Nd from used magnets\cite{Tanaka:book2013,Binnemans:JCP2013}, as one of provisions for stock. 

The recycle method can be divided into two main classes: wet and dry processes.
Associated with the development of ore dressing technologies, the wet process has already been put to practical use in the recycle of sludge of in-plant scrap.
The scrap is dissolved in acid such as HCl, HNO$_{3}$ and H$_{2}$SO$_{4}$.
After the filtration of insoluble materials mainly containing Fe, acid solution is reacted with oxalic or carbonic acid to form the precipitate containing Nd element\cite{Uchida:unpat1983}.
The calcined precipitate becomes Nd oxide, which can be returned to the initial manufacture process of Nd-magnet.
The oxidation of Nd-Fe-B magnet as the pretreatment improves the selectivity between Nd and Fe, however the recovery ratio of Nd is rather low\cite{Santoku:Jappat,Santoku:unpatH9}.
Nearly 100\% recovery is achieved when the acid solution is heated at 180 $^{\circ}$C\cite{JO:Jappat}.

As for the dry process, Takeda et al. have demonstrated the Nd-recovery by employing Mg acting as an extraction medium, which forms a low-viscosity liquid-alloy with Nd\cite{Takeda:JACP2006}.
Uda et al. have investigated the rare-earth metal separation by a selective reduction and a distillation\cite{Uda:Science2000}.
The large difference of vapor pressure between RECl$_{2}$ and RECl$_{3}$ (RE:rare earth) is skillfully utilized and the selection efficiency is highly improved.
Matsumiya et al. have reported that the electrodeposition in ionic liquid of environmental-friendly can separate iron group metal and recover Nd metal\cite{Matsumiya:JJIM2011}. 
Recently a mixture with flux FeO$\cdot$B$_{2}$O$_{3}$ is a promising dry process for high purity and high extraction ratio of rare-earth oxide\cite{Bian:JSM2015}.  
Although many attempts based on the dry process have been proposed so far, they are still in the fundamental stage.

Returning to the wet process, it is desired to increase the selectivity between Nd and Fe at room temperature.
We have found that the corrosion process leads to the high selectivity at room temperature and nearly 100\% recovery of Nd.
Our method is compatible with the present recovery method in the in-plant, because the oxidation process is merely replaced with the corrosion one.
In this report, we have investigated the Nd-recovery ratio in the recovery method using corrosion as the pretreatment.
The ratio is compared with that obtained by the conventional method in the in-plant.

\section{Experimental method}

\begin{figure}
\begin{center}
\includegraphics[width=6cm]{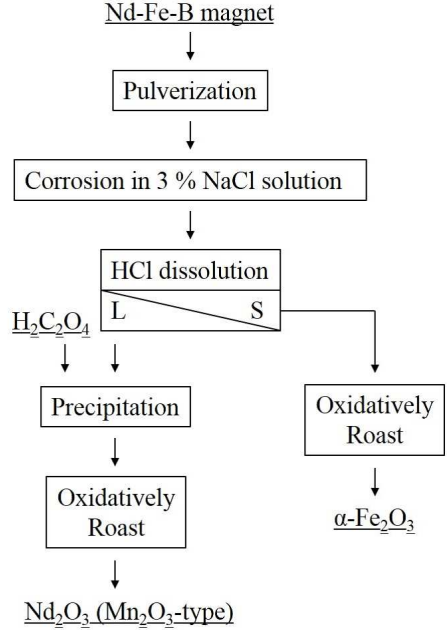}
\end{center}
\caption{Process flow of proposed Nd-recovery method.}
\label{f1}
\end{figure}

Figure 1 shows the process flow of proposed method.
A demagnetized and coarsely-ground commercial Nd-Fe-B magnet (Niroku Seisakusyo, NK094), weighting approximately 0.5 g, was immersed into 3 \% NaCl solution for 1 week.
The elemental component according to the manufacturer is Nd:Fe:B:otherwise (Dy et al.) $=$ 28:66:1:5 in wt\%.
An air pump provided constant airflow to the solution in order to accelerate the corrosion. 
The corroded sample was dissolved, at room temperature, into HCl solution ranging from 0.1 mol/L to 0.3 mol/L.
The insoluble after the dissolution was oxidatively roasted at 800 $^{\circ}$C for 5 h by an electric furnace to obtain $\alpha$-Fe$_{2}$O$_{3}$.
After removal of the insoluble, the solution was reacted with 0.02 mol/L oxalic acid.
The precipitate after the reaction was also oxidatively roasted at 800 $^{\circ}$C for 5 h to obtain cubic Mn$_{2}$O$_{3}$-type Nd$_{2}$O$_{3}$ (c-Nd$_{2}$O$_{3}$).
The samples were evaluated using the powder X-ray diffraction (XRD) patterns (Cu-K$\alpha$ radiation).
We employed Inductively Coupled Plasma (ICP) spectroscopy to analyze the composition of c-Nd$_{2}$O$_{3}$ and its purity.

\begin{figure}
\begin{center}
\includegraphics[width=8cm]{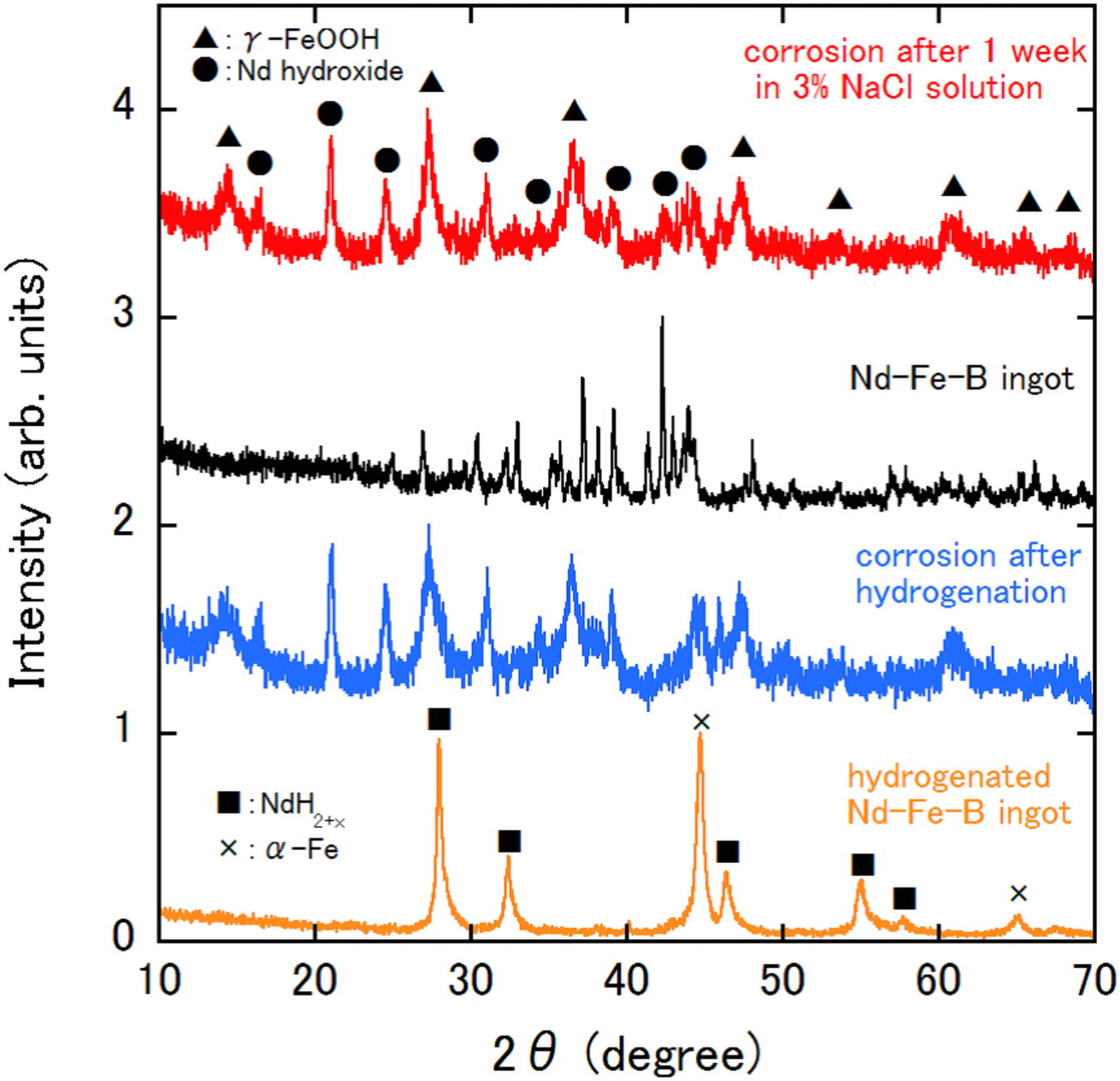}
\end{center}
\caption{XRD patterns of sample corroded after 1 week, Nd-Fe-B ingot, sample corroded after hydrogenation process and hydrogenated Nd-magnet. The origin of each pattern is shifted by an integer value for clarity.}
\label{f2}
\end{figure}

\section{Results and discussion}
The XRD pattern of corroded sample after 1 week is shown in Fig.\ 2, where the XRD pattern of Nd-Fe-B ingot is also displayed.
The main composition of Nd-Fe-B ingot is Nd$_{2}$Fe$_{14}$B with additional minor phase of NdFe$_{4}$B$_{4}$.
The XRD pattern of Nd-Fe-B ingot completely disappears in the corroded sample, partially containing the XRD pattern of $\gamma$-FeOOH denoted by the filled triangles.
The rest of diffraction peaks in the corroded sample would be assigned to a Nd compound (filled circles).
To confirm the expectation, we have performed the corrosion of Nd-Fe-B ingot hydrogenated at 600$^{\circ}$C for 12 h under a high pressure of hydrogen.
The hydrogenated sample\cite{Kataoka:RIP2015} shows the decomposion into Nd hydride (NdH$_{2+x}$) and $\alpha$-Fe (see the bottom pattern of Fig.\ 2).
Therefore the corrosion process independently occurs in each compound.
The XRD pattern of corroded sample after the hydrogenation almost coincides with that of directly corroded Nd-Fe-B ingot.
Considering that, in the corrosion process, $\alpha$-Fe transforms into the Fe hydroxide of $\gamma$-FeOOH, a Nd hydroxide is probably responsible for the rest of the diffraction peaks in corroded sample.
It should be noted that the NaCl concentration is not optimized.
We merely suppose the sea water, which is abundant resource.
The preliminary result using more concentrate NaCl solution (10 \%) also leads to the corroded state.
The further study of another NaCl concentration is needed in the future. 

\begin{figure}
\begin{center}
\includegraphics[width=10cm]{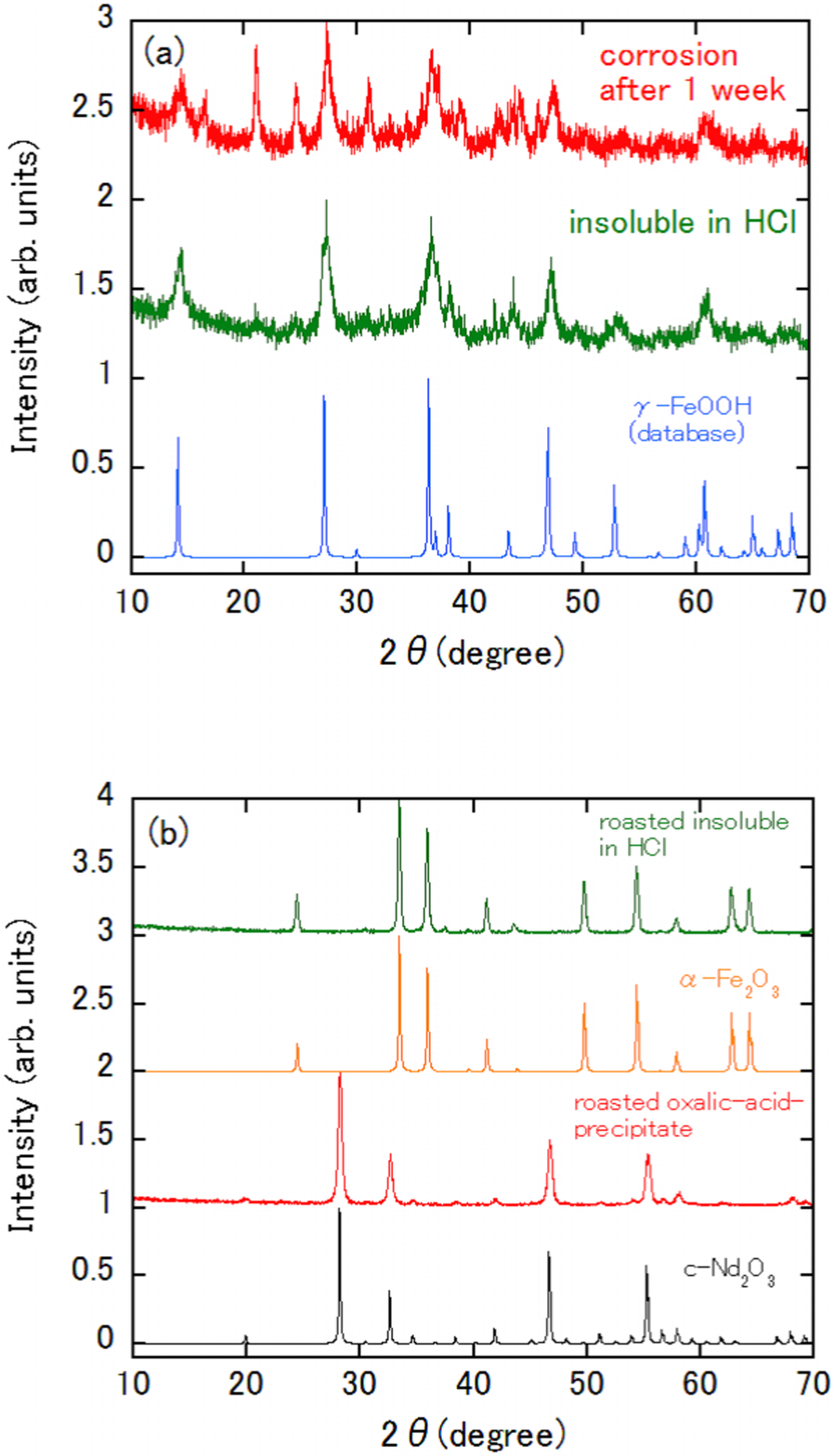}
\end{center}
\caption{(a) XRD patterns of sample corroded after 1week, insoluble material in 0.1 mol/L HCl dissolution after 30 min and $\gamma$-FeOOH (database). (b) XRD patterns of insoluble and oxalic acid precipitate, in HCl solution, after oxidatively roast. The database patterns of $\alpha$-Fe$_{2}$O$_{3}$ and c-Nd$_{2}$O$_{3}$ are also shown. The concentration of HCl is 0.2 mol/L and the immersion time is 2 h. The origin of each pattern in (a) or (b) is shifted by an integer value for clarity.}
\label{f3}
\end{figure}

\begin{figure}
\begin{center}
\includegraphics[width=10cm]{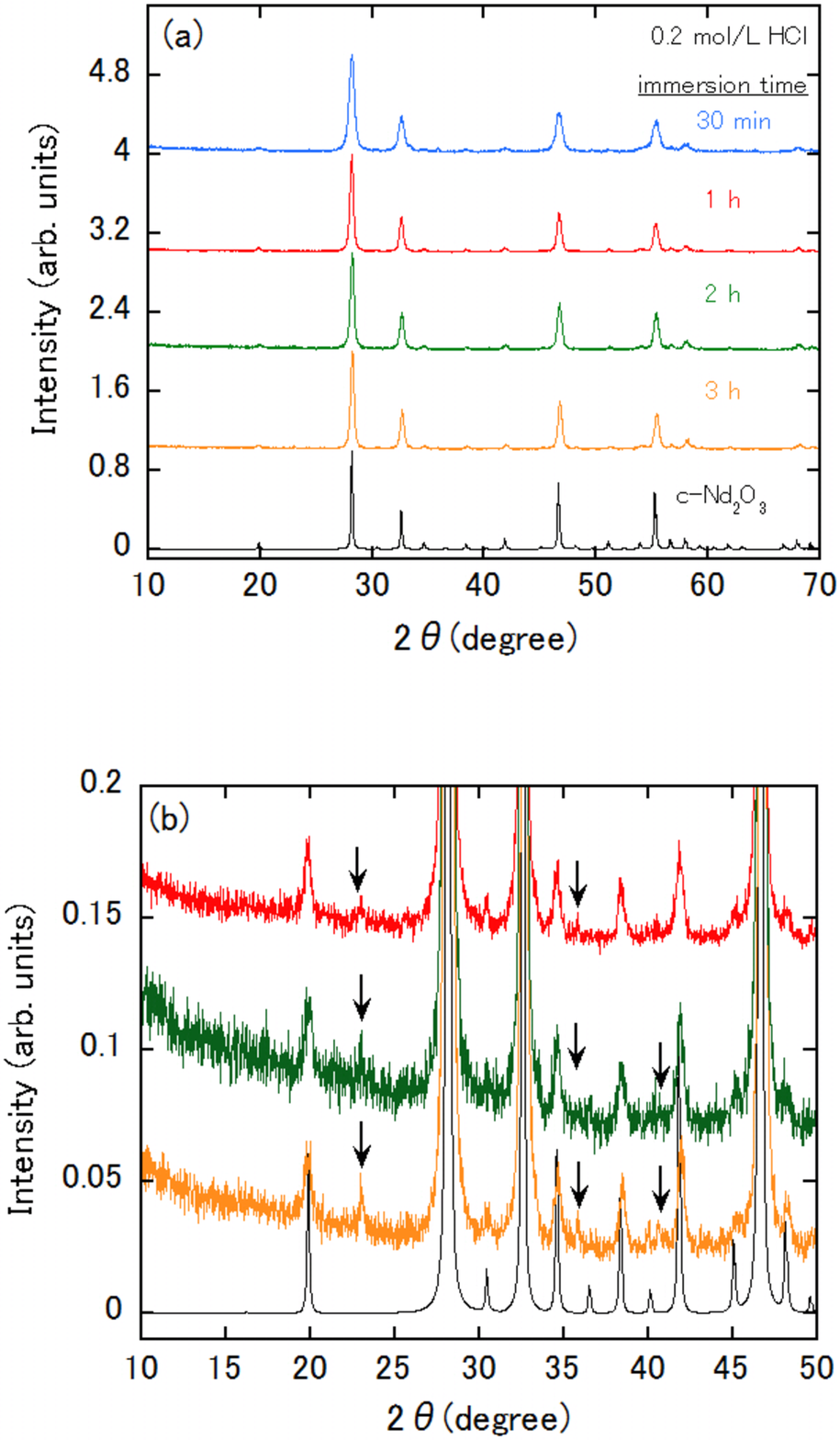}
\end{center}
\caption{(a) XRD patterns of recovered c-Nd$_{2}$O$_{3}$ through 0.2 mol/L HCl dissolution with immersion-time 30 min, 1 h, 2 h and 3 h. The database pattern of c-Nd$_{2}$O$_{3}$ is also given. The origin of each pattern is shifted by an integer value for clarity. (b) Expanded view of (a). The arrows indicate impurity peaks. The color corresponds to that in (a).}
\label{f4}
\end{figure}

We checked the XRD pattern of residue left after HCl dissolution of the corroded sample as shown in Fig.\ 3(a).
The diffraction peaks are in good agreement with the XRD pattern of $\gamma$-FeOOH.
Oxidatively-roasted $\gamma$-FeOOH exhibits the XRD pattern of $\alpha$-Fe$_{2}$O$_{3}$ (see Fig.\ 3(b)).
Figure 3(a) indicates that the Nd hydroxide would be selectively dissolved into HCl solution, in which Nd ions are formed.
The oxalic acid precipitation was carried out to recover Nd.
The precipitate has been oxidatively roasted and evaluated by XRD pattern, which is displayed in Fig.\ 3(b) with the database pattern of c-Nd$_{2}$O$_{3}$.
The XRD patterns are well matched between the roasted precipitate and c-Nd$_{2}$O$_{3}$, suggesting the successful recovery of Nd in the form of Nd-oxide.

Figure 4(a) shows the example of XRD patterns of oxidatively-roasted oxalic-acid-precipitate, obtained after 0.2 mol/L HCl dissolution with the immersion time varying from 30 min to 3 h.
In all cases, the XRD patterns matching well with that of c-Nd$_{2}$O$_{3}$ (database) suggest the successful recovery of Nd.
The expanded view given in Fig.\ 4(b) has revealed the impurity phases indicated by arrows.
The intensity of impurity peaks gradually grows with increasing time of immersion into HCl solution.
The results through 0.1 or 0.3 mol/L HCl dissolution exhibit the same tendency as Fig.\ 4(b).

\begin{figure}
\begin{center}
\includegraphics[width=8cm]{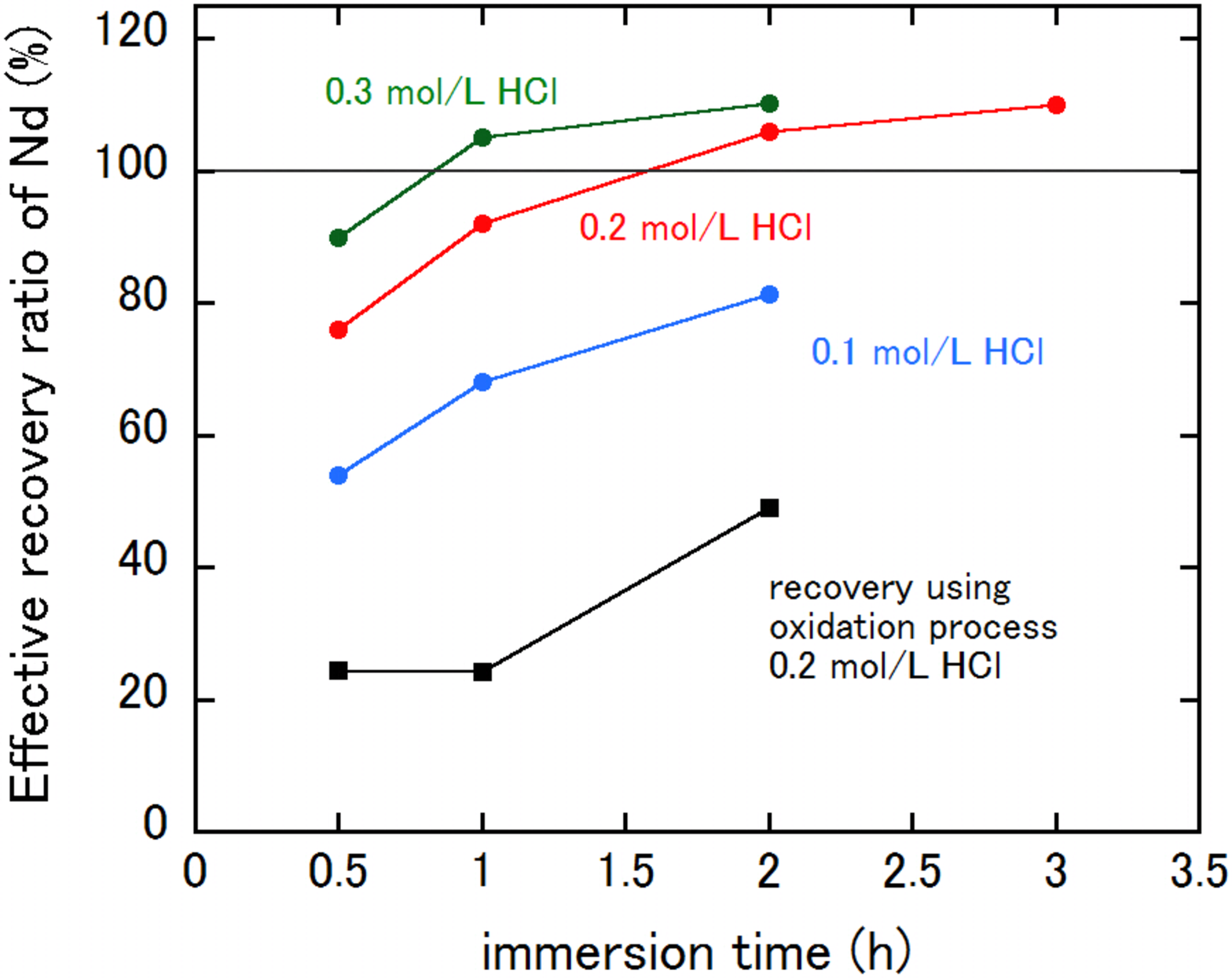}
\end{center}
\caption{Immersion-time dependences of effective recovery-ratio of Nd through 0.1, 0.2 and 0.3 mol/L HCl dissolution. The effective recovery-ratio obtained by the conventional recovery method using the oxidation process is also shown.}
\label{f5}
\end{figure}

\begin{table}
\caption{Compositions of c-Nd$_{2}$O$_{3}$ and $\alpha$-Fe$_{2}$O$_{3}$ analyzed by ICP spectroscopy. ND means not detected.}
\label{t1}
\begin{tabular}{ccc}
\hline
composition (at\%) & c-Nd$_{2}$O$_{3}$ & $\alpha$-Fe$_{2}$O$_{3}$ \\
\hline
Nd & 97.2 & 4.0 \\
Fe & 2.8 & 95.8 \\
B & ND & 0.2 \\
\hline
\end{tabular}
\end{table}

Figure 5 is the immersion-time dependences of effective recovery-ratio $R$ of Nd through 0.1, 0.2 and 0.3 mol/L HCl dissolution.
$R$ was calculated by the equation,
\begin{equation}
R=\frac{\rm{Mass\quad of\quad Nd\quad in \quad {Nd_{2}O_{3}}}}{\rm{Mass\quad of\quad Nd\quad in \quad Nd-magnet}}\times 100.
\label{equ:RT}
\end{equation}
For each HCl concentration, $R$ increases with prolonged immersion-time, and approximately saturates, however exceeds 100\% at some conditions.
The excess above 100\% would be due to the existence of impurity phases in c-Nd$_{2}$O$_{3}$.
The compositions of c-Nd$_{2}$O$_{3}$ and $\alpha$-Fe$_{2}$O$_{3}$ analyzed by the ICP spectroscopy are summarized in Table I.
These samples were obtained after 0.2 mol/L HCl dissolution with the immersion time of 2 h.
Since the ICP spectroscopy cannot detect O element, we do not note the composition of O.
We can estimate the net recovery-ratio of Nd by using the fact that $\alpha$-Fe$_{2}$O$_{3}$ contains 4 at\% Nd.
Subtracting the mass of Nd left in $\alpha$-Fe$_{2}$O$_{3}$ from that in starting Nd-Fe-B ingot, the recovery ratio is calibrated to be 93\%.
Nd site in c-Nd$_{2}$O$_{3}$ would be slightly replaced with Fe, as indicated by Nd$_{1.944}$Fe$_{0.056}$O$_{3}$.
Taking into account the ratio between the effective and net recovery ratios, and the composition of Nd$_{1.944}$Fe$_{0.056}$O$_{3}$, the purity of c-Nd$_{2}$O$_{3}$ is determined to be 85\%.
Assuming that the other $R$s follow the same calibration coefficient, the highest net-recovery-ratio in Fig.\ 5 is 97\%.
The 100\% recovery ratio would be achieved with increasing HCl concentration and/or immersion time, however the amount of impurity phases would be increased.
We have to introduce a sophisticated step such as pH adjustment in the oxalic acid precipitation to reduce the incorporation of impurity phases.

\begin{figure}
\begin{center}
\includegraphics[width=10cm]{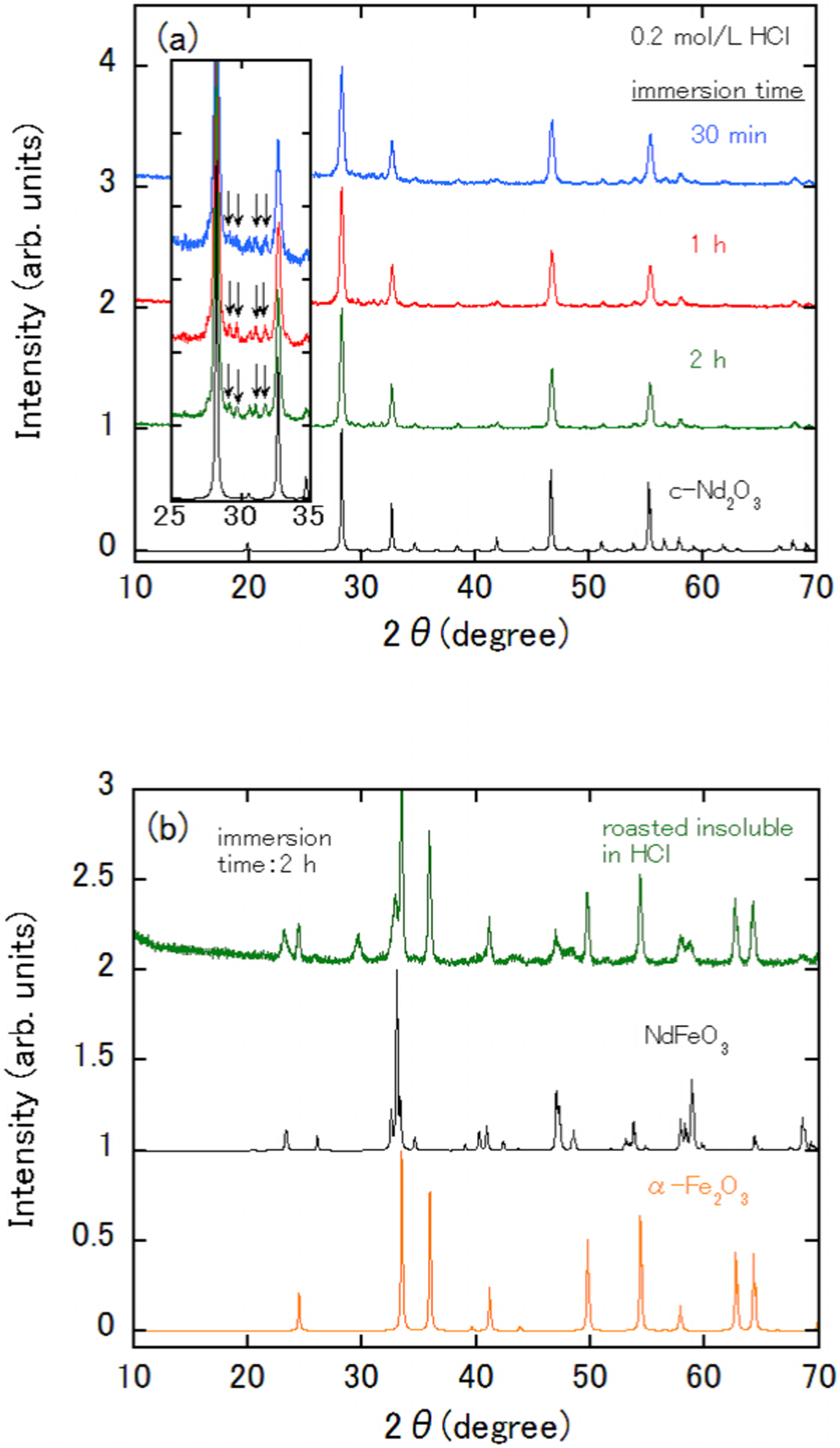}
\end{center}
\caption{(a) XRD patterns of recovered c-Nd$_{2}$O$_{3}$ by conventional recovery method, using oxidation of Nd-Fe-B magnet, with varying immersion-time (0.2 mol/L HCl) from 30 min to 2 h. The inset shows the expanded view with 2$\theta$ between 25$^{\circ}$ and 35$^{\circ}$. The arrows indicate impurity phases. (b) XRD patterns of oxidatively-roasted residue left after 0.2 mol/L HCl dissolution. The database patterns of NdFeO$_{3}$ and $\alpha$-Fe$_{2}$O$_{3}$ are also shown. The origin of each pattern in (a) or (b) is shifted by an integer value for clarity.}
\label{f6}
\end{figure}

We revisited the conventional Nd-recovery-method using oxidation process.
The coarsely-ground commercial Nd-Fe-B magnet after the oxidization at 600 $^{\circ}$C for 5 h was dissolved into 0.2 mol/L HCl solution, at room temperature, with varying immersion-time from 30 min to 2 h.
The oxidized Nd-Fe-B decomposes into Nd$_{2}$O$_{3}$ and $\alpha$-Fe$_{2}$O$_{3}$.
In Fig.\ 6(a), the XRD pattern of recovered sample after the oxalic acid precipitation for each condition is displayed.
Although c-Nd$_{2}$O$_{3}$ is recovered, several impurity peaks, which are different from those in Fig.\ 4(b), appear (see arrows in the inset of Fig.\ 6(a)).
Figure 6(b) shows the XRD pattern of oxidatively-roasted insoluble in HCl, and also the database patterns of NdFeO$_{3}$ and $\alpha$-Fe$_{2}$O$_{3}$, the superposition of which reproduces the experimental data.
The existence of NdFeO$_{3}$ means that Nd stays in the insoluble in HCl solution.
The immersion-time dependence of $R$ for the method based on the oxidation is added in Fig.\ 5.
The $R$ is much lower than that for the proposed method, reflecting the observation of NdFeO$_{3}$ in the insoluble in HCl solution.

Even though the preparation of $\alpha$-Fe$_{2}$O$_{3}$ by the oxidation process prevents the incorporation of Fe in the recovered material\cite{Santoku:Jappat,Santoku:unpatH9}, Nd-recovery ratio is rather low.
According to the patent\cite{JO:Jappat}, the high-temperature HCl-dissolution improves the recovery ratio near to 100\%.
This means that the solubility of Nd$_{2}$O$_{3}$ is low at room temperature.
On the other hand, our proposed method, where the oxidation process is replaced with the corrosion one, have no need of temperature-rise to obtain enough solubility of Nd into HCl solution.
After the corrosion, Nd-Fe-B magnet decomposes into $\gamma$-FeOOH and Nd hydroxide.
The preparation of these compounds is the key factor to realize the high selectivity between Nd and Fe even at room temperature.

There is not yet a study, carrying out the proposed Nd recovery process for used magnets.
The elemental component of Nd-magnet depends on the kind of products, and we speculate that the process condition should be finely tuned for each composition.
Therefore the sort of the kind of products might be necessary.
In addition, the study for Nd-magnet with elemental component other than that in this study is further needed.
We note here that the corrosion process allows the wide applicable range of Nd-magnet condition; for example, the proposed process is highly compatible with a magnet corroded due to peeled coating in using for many years.
Furthermore our process may also allow a direct corrosion of a roughly disassembled product, and the recycle cost can be reduced.

\section{Summary}
We have demonstrated the Nd-recovery from Nd-Fe-B magnets by improving the in-plant recycle method, based on the wet process using HCl solution and oxalic acid precipitation.
The pretreatment of corrosion plays the important role for the high selectivity between Nd and Fe even at room temperature.
After the oxalic acid precipitation, c-Nd$_{2}$O$_{3}$ is obtained.
The recovery ratio of Nd reaches to 97\%, which is much higher than that obtained by the conventional recovery method using oxidation.
We emphasize that our proposed method is compatible with the present in-plant recycling of sludge.

\end{document}